%Paper: 9112056
%From: WITTEN%IASSNS.BITNET@pucc.PRINCETON.EDU
%Date: Thu, 19 Dec 91 15:52 EST

\input harvmac

\def\etag#1{\eqnn#1\eqno#1}
\def\subsubsec#1{{\medbreak\smallskip\noindent{\it #1 } }}

\def\inbar{\,\vrule height1.5ex width.4pt depth0pt}
\def\IC{\relax{\hbox{$\inbar\kern-.3em{\rm C}$}}}
\def\IR{\relax{\rm I\kern-.18em R}}
\font\cmss=cmss10 \font\cmsss=cmss10 at 7pt
\def\IZ{\relax\ifmmode\mathchoice
{\hbox{\cmss Z\kern-.4em Z}}{\hbox{\cmss Z\kern-.4em Z}}
{\lower.9pt\hbox{\cmsss Z\kern-.4em Z}}
{\lower1.2pt\hbox{\cmsss Z\kern-.4em Z}}\else{\cmss Z\kern-.4em Z}\fi}

%\ref\froh
%{J. Ambjorn, B. Durhuus, and J. Frohlich, {\it
%Diseases of Triangulated Random
%Surface Models, and Possible Cures}
%Nucl. Phys. {B257[FS14]} (1985) 433; J. Ambjorn, B. Durhuus, J. Frohlich,
%and P. Orland, {\it  The Apppearance Of Critical Dimensions in
%Regulated String Theories,} Nucl. Phys. {\bf B270[FS16]} (1986) 457.}

\lref\aspinwall{P. S. Aspinwall and D. R. Morrison, 
``Topological Field Theory And
Rational Curves,'' Oxford and Duke preprints OUTP-91-32P, DUK-M-91-12.}

\lref\atiyah{M. F. Atiyah and L. Jeffrey, 
``Topological Lagrangians And Cohomology,''
J. Geom. Phys. {\bf 7} (1990) 119.}

\lref\baulieu{L. Baulieu and I. M. Singer, 
``The Topological Sigma Model,'' Commun.
Math. Phys. {\bf 125} (1989) 227.}

\lref\blok{B. Blok and A. Varchenko, 
``Topological Conformal Field Theories
And The Flat Coordinates,'' IASSNS 91-5.}

\lref\candelas{P. Candelas, P. Green, 
L. Parke, and X. de la Ossa,  ``A Pair Of
Calabi-Yau Manifolds As An Exactly Soluble Superconformal Field Theory,''
Nucl. Phys. {\bf B359} (1991) 21.}

\lref\dijkgraaf{R. Dijkgraaf, E. Verlinde, and 
H. Verlinde, ``Topological Strings In $D<1$,''
IASSNS-HEP-90/71.}

\lref\dine{M. Dine, N. Seiberg, X.-G. Wen, and E. Witten,  
``Nonperturbative Effects
On The String World Sheet I,II,'' Nucl. Phys. {\bf B278} (1986) 769,
{\bf 289} (1987) 319.}

\lref\eguchi{T. Eguchi and S.-K. Yang, 
``$N=2$ Superconformal Models As Topological
Field Theories,'' Tokyo preprint UT-564 (1990).}

\lref\greene{B. Greene and R. Plesser, 
``Duality In Calabi-Yau Moduli Space,'' Nucl.
Phys. {\bf B338} (1990) 15.}

\lref\gromov{M. Gromov, 
``Pseudoholomorphic Curves In Symplectic Manifolds,''
Invent. Math. {\bf 82} (1985) 307.}

\lref\gromov{M. Gromov and M. A. Shubin, 
``The Riemann-Roch Theorem For General Elliptic
Operators,'' preprint.}

\lref\landweber{P. S. Landweber, ed., 
``Elliptic Curves And Modular Forms in Algebraic
Topology,'' Lecture Notes In Mathematics 1326 (Springer-Verlag, 1988).}

\lref\lerche{W. Lerche, C. Vafa, and N. P. Warner, ``Chiral Rings in $N=2$
Superconformal Theories,'' Nucl. Phys. {\bf B324} (1989) 427.}

\lref\morrison{D. R.  Morrison,  ``Mirror Symmetry And Rational Curves On
Quintic Threefolds: A Guide For Mathematicians,'' Duke preprint DUK-M-91-01. }

\lref\saito{K. Saito, ``Period Mapping Associated To A Primitive Form,'' Publ.
RIMS, Kyoto Univ. {\bf 19} (1983) 1231.}

\lref\vvafa{C. Vafa. ``Topological Landau-Ginzberg Models,'' Harvard preprint,
(1990).}

\lref\vafa{C. Vafa, ``Topological Mirrors And Quantum Rings,'' contribution
to this volume.}

\lref\vvvafa{S. Cecotti and C. Vafa, ``Topological Anti-Topological Fusion,''
Harvard University preprint (1991).}

\lref\oldwitten{E. Witten, 
``Topological Sigma Models,'' Commun. Math. Phys. {\bf 118}
(1988) 411. }

\lref\witten{E. Witten, ``The $N$ Matrix Model and Gauged WZW Models,''
IAS preprint, to appear in Nucl. Phys. B.}

\line{\hfill IASSNS-HEP-91/83}
\line{\hfill December,  1991}
\bigskip\medskip
\vskip .5cm
\centerline{{MIRROR MANIFOLDS}}
\medskip
\centerline{{AND TOPOLOGICAL FIELD THEORY}}
\bigskip\medskip
\centerline{{\authorfont Edward Witten}\footnote*{Research supported in part
by NSF Grant 91-06210.}}
\centerline{\it School of Natural Sciences,}
\centerline{\it Institute for Advanced Study,}
\centerline{\it Olden Lane,}
\centerline{\it Princeton, N.J. 08540}
\bigskip\bigskip
\centerline{ABSTRACT}
\bigskip
\noindent
These notes are devoted to sketching  how some of the standard
facts  relevant to
mirror symmetry and its applications
can be naturally understood in the context of topological field theory.
If $X$ is a Calabi-Yau manifold, the usual nonlinear sigma model
governing maps of a Riemann surface $\Sigma$ to $X$ can be twisted
in two ways to give topological field theories, which I call the
$A$ model and the $B$ model.  Mirror symmetry relates the $A$ model
(of one Calabi-Yau manifold) to the $B$ model (of its mirror).
The correlation functions of the $A$ and $B$ models can be computed,
respectively, by counting rational curves and by calculating periods
of differential forms.  This can be proved as a consequence of a reduction to
weak coupling (as in \S3-4 of these notes) or by a sort of fixed
point theorem for the Feynman path integral
(see \S5).  The correlation functions of the twisted
models coincide, as explained in \S6,
with certain matrix elements of the physical, untwisted model
-- namely those that determine the superpotential.
The conventional moduli spaces of sigma models can be thickened,
in the context of topological field theory, to extended moduli
spaces, indicated in \S7, which are probably the natural framework
for understanding the still mysterious
``mirror map'' between moduli spaces.

\vfill\eject

\newsec{Introduction}

%$$ I={1\over 2\pi}\int \sqrt g~R  \etag\hilbact$$
%[\poly], noncritical two dimensional gravity is related
%[\gerv,\curt] seeking
%\def\cgn{\overline{{\cal M}}_{g,n}}
%\def\om{\overline{{\cal M}}}
%$$ \sum_{i=1}^n~d_i=3g-3+n.  \etag\dimcon$$
%\foot{In [\witt], I worked with $\sigma_d=d!\tau_d$, in order
%one sees that the path integral in \lolf\ is the generating function of the
%\subsubsec{Consequences Of The Conjecture}

%\subsec{2b. Evidence For The Conjecture}

%\FIG\doubline{A convenient notation
%$$\eqalign{\int (dM)\exp\left(-\Tr(M^2/2)\right)& M^{i_i}{}_{j_1}

The purpose of these notes is to explain aspects of the mirror
manifold problem that can be naturally understood in the context
of topological field theories.  (For other aspects of the problem,
and detailed references, the reader should consult other articles in this
volume.)  Most of what I will say can be found
in the existing literature, but
to isolate the facts most relevant to mirror symmetry may be useful.
The points that I want to explain are as follows.

First, we will consider the standard supersymmetric nonlinear sigma model
in two dimensions, governing maps of a Riemann surface $\Sigma$ to
a target space which for our purposes will be a Kahler manifold $X$
of $c_1=0$.  We will see that there are two different topological field
theories that can be made by twisting the standard sigma model.  There
are not standard names for these topological field theories; I will call
them the $A$ theory and the $B$ theory, or $A(X)$ and $B(X)$ when I want
to specify the choice of $X$.  The $A$ theory has been studied in detail
in [\oldwitten], where many facts sketched below are explained;
the $B$ theory has been studied less intensively.  (The $A$ and $B$ theories
were discussed qualitatively, in a general context of $N=2$
superconformal field theories, by Vafa in his lecture at this meeting
[\vafa].)

Unlike the ordinary supersymmetric nonlinear sigma model, the twisted
models are ``soluble'' in the sense that the problem of computing
all the physical observables can be reduced to classical questions in
geometry.  This is done by a sort of fixed point theorem in field space,
which gives the following results: the correlation functions of the $A$
theory are determined by counting holomorphic maps $\Sigma\to X$
obeying various conditions; the correlation functions of the $B$
theory can be computed by calculating periods of classical differential forms.
(More generally, ``counting'' rational curves must be replaced by
computing the Euler class of the vector bundle of antighost zero modes,
as I have explained
elsewhere [\witten, \S3.3]; this has been
implemented in the mirror manifold problem
by Aspinwall and Morrison [\aspinwall].)

For the most direct physical applications one is not interested
in the twisted $A$ or $B$ theories, but in the original, ``physical''
sigma model.  However, there is one very important case in which physical
observables coincide with observables of the $A$ and $B$ theories.
This happens for the Yukawa couplings, which are certain quantities
that one wishes to compute for $\Sigma$ of genus zero.\foot{The 
restriction to genus zero appears in two ways:
the superpotential that determines the Yukawa couplings has no higher
genus corrections because of nonrenormalization theorems [\dine];
alternatively, in relating the superpotential to an observable of
the twisted model, we will require in \S6
that the canonical bundle of $\Sigma$
with two points deleted be trivial, which is of course true only in genus
zero.}
In particular, the $\overline {\bf 27}^3$ and $
{\bf 27}^3$ Yukawa couplings of superstring models coincide with certain
observables of the $A$ and $B$ theories, respectively.  These Yukawa
couplings are therefore determined by
the fixed point theorem mentioned in the last paragraph;
from this we recover
results that were originally found
by more detailed arguments, such as the fact that the ${\bf 27}^3$
Yukawa couplings have no quantum corrections, and the instanton sum formula
[\dine] for the $\overline {\bf 27}^3$ Yukawa couplings.

In \S2 we recall the definition of the standard sigma model.
The twisted $A$ and $B$ models are described in \S3 and \S4,
along with an explanation of their main properties, including
the reduction to instanton moduli spaces and to constant maps in
the $A$ and $B$ models respectively.  This latter reduction is
reinterpreted as a sort of fixed point theorem in \S5.
In \S6, I explain why certain observables of the standard
physical sigma model coincide with observables of the twisted models.
This occurs whenever the canonical bundle of
the Riemann surface becomes trivial after deleting the points at which
fermion vertex operators have been inserted -- in practice, mainly
for amplitudes in genus zero with precisely two fermions.
In \S7 -- the only part of these notes containing some novelty --
I look a little more closely at the $A$ and $B$ models
and describe the full families of topological field theories of which
they are part.
I strongly suspect that many properties of mirror symmetry that are not
now well understood, including the structure of the mirror map between
the parameter spaces, can be better understood in looking at the full
topological families.

%A ``mirror pair'' of Calabi-Yau manifolds is a pair, say $X$
%and $Y$, with the property that the nonlinear sigma models with target
%spaces $X$ and $Y$ are isomorphic, by an isomorphism that reverses
%the sign of one of the $U(1)$ charges.
%The latter reversal exchanges the $A$ and $B$ theories, so mirror
%symmetry induces isomorphisms $A(X)\cong B(Y)$ and $A(Y)\cong B(X)$.
%Thus, rational curves on $X$ of given degree
%can be counted by calculating periods of
%differential forms on $Y$, and vice-versa.

%Greene and Plesser identified some explicit examples of
%mirror pairs $X,Y$ [\greene]; and for one of those pairs, Candelas, Green,
%Parke, and de la Ossa [\candelas]
%computed the relevant periods of differential
%forms on $Y$, ultimately getting detailed formulas for the number of
%rational curves on $X$ of various degrees.
%In carrying this out in full, it is essential to understand the mirror
%map, that is, the map induced by mirror symmetry
%between the relevant $X$ and $Y$ moduli spaces.
%Candelas et. al. determined this map in their example by a natural
%ansatz (expressed most simply by Morrison [\morrison])
%which has not been justified
%even at a physical level of rigor.  I will conclude these
%%notes by attempting to use properties of topological field theories
%to identify the mirror map.  This is not yet fully understood, but I hope
%to point out some of the relevant issues.

Since ``elliptic genera'' (see [\landweber])
arise in the same supersymmetric nonlinear
sigma models that are the basis for the study of mirror manifolds,
I will also along the way make a few observations about them.
In particular we will note the easy fact that if $X,Y$ are a mirror pair,
they have the same elliptic genus.  This is trivial
in complex dimension three (since the elliptic genus of a three
dimensional Calabi-Yau manifold is zero), but becomes interesting
in higher dimension.

Perhaps I should emphasize that if $X$ and $Y$ are
a mirror pair, then mirror symmetry relates {\it all} observables
of the sigma model on $X$ to corresponding observables on $Y$ -- and not
just the few observables that can be related naturally to the
twisted models and thus to topological field theory.
The literature on mirror symmetry has focussed
on these particular observables, which of course are also the ones
we will be studying here, because of their phenomenological
importance, and because, since the $B$ model is soluble classically,
the relation given by mirror symmetry between observables of the $A(X)$ model
and observables of the $B(Y)$ model is particularly useful.

\newsec{Preliminaries}

To begin with, we recall the standard supersymmetric nonlinear
sigma model in two dimensions.\foot{This discussion will be at the
classical
level, and we will not worry about the anomalies that arise and spoil
some assertions if the target space is not a Calabi-Yau manifold.}
It governs maps $\Phi:\Sigma\to X$,
with $\Sigma$ being a Riemann surface and $X$ a Riemannian manifold of
metric $g$.  If we pick local coordinates $z,\overline z$ on $\Sigma$
and $\phi^I$ on $X$, then $\Phi$ can be described locally via functions
$ \phi^I(z,\bar z)$.  Let $K$ and $\overline K$ be the canonical and
anti-canonical line bundles of $\Sigma$ (the bundles of one forms of types
$(1,0)$ and $(0,1)$, respectively), and let $K^{1/2}$ and $\overline K^{1/2}$
be square roots of these.  Let $TX$ be the complexified tangent bundle
of $X$.  The fermi fields of the model are $\psi_+^I$, a section
of $K^{1/2}\otimes \Phi^*(TX)$, and $\psi_-^I$, a section of
$\overline K^{1/2}\otimes \Phi^*(TX)$.  The Lagrangian is\foot{Here 
$d^2z$ is the measure $-idz\wedge d\bar z$.  Thus if $a$ and
$b$ are one forms, $\int a\wedge b = i\int d^2z \left(a_zb_{\bar z}
-a_{\bar z}b_z\right)$.  The Hodge $\star$ operator is defined by
$\star dz=idz$, $\star d\bar z=-id\bar z$.}
$$\eqalign{L=&2
t\int d^2z\left({1\over 2}g_{IJ}(\Phi)\partial_z\phi^I\partial_{\bar z}
\phi^J +{i\over 2}g_{IJ}\psi_-^ID_z\psi_-^J
\right.\cr 
&\quad\quad\quad\quad+\left.{i\over 2}g_{IJ}\psi_+^I
D_{\overline z}\psi_+^J+{1\over 4}R_{IJKL}\psi_+^I\psi_+^J\psi_-^K\psi_-^L
\right).\cr} \etag\albo$$
Here $t$ is a coupling constant,
and $R_{IJKL}$ is the Riemann tensor of $X$.
$D_{\bar z}$ is the
$\bar\partial$ operator on $K^{1/2}\otimes
\Phi^*(TX)$ constructed using
the pullback of the Levi-Civita connection on $TX$.
In formulas (using
a local holomorphic trivialization of $K^{1/2}$),
$$D_{\bar z}\psi_+^I={\partial\over\partial \overline z}\psi_+^I
+{\partial \phi^J\over\partial\overline z}\Gamma^I_{JK}\psi_+^K, \etag\balfo$$
with $\Gamma^I_{JK}$ the affine connection of $X$.
Similarly $D_z$ is the $\partial$ operator on
$\overline K^{1/2}\otimes \Phi^*(TX)$.

The supersymmetries of the model are generated by infinitesimal transformations
$$\eqalign{
  \delta \Phi^I & =i\epsilon_-\psi_+^I+ i\epsilon_+\psi_-^I \cr
  \delta \psi_+^I & = -\epsilon_-\partial_z\phi^I-i\epsilon_+\psi_-^K\Gamma
 ^I_{KM}\psi_+^M\cr
  \delta \psi_-^I & = -\epsilon_+\partial_{\overline z}\phi^I-i\epsilon_-
\psi_+^K\Gamma^I_{KM}\psi_-^M, \cr }   \etag\congo$$
where $\epsilon_-$ is a holomorphic section of $K^{-1/2}$, and $\epsilon_+$
is an antiholomorphic section of $\overline K^{-1/2}$.
The formulas for the Lagrangian \albo\ and the transformation laws
\congo\ have a natural interpretation upon formulating the model
in superspace, that is in terms of maps of a super-Riemann surface to $X$.
I will not discuss this here; for references see Rocek's lecture in this
volume.

As a small digression, let me note that
usually, in discussing superconformal symmetry, one considers
$\epsilon$'s that are defined locally (say in a neighborhood of a circle
$C\subset \Sigma$ if one is studying the Hilbert spaces obtained by
quantization on $C$).
One might wonder what happens if we can find global $\epsilon$'s.
On a Riemann surface of genus $g>1$, this will not occur as there
are no holomorphic sections of $K^{-1/2}$.
In genus one $K$ is trivial; if we pick $K^{1/2}$ to be trivial,
then there is a one dimensional space of global $\epsilon_-$'s.
On the other hand, if we pick $\overline K^{1/2}$ to be a non-trivial
line bundle (of order two)
then globally $\epsilon_+$ must vanish.  With the techniques that
we will use below, one can readily use the symmetry generated by $\epsilon_-$
to prove that the partition function of the sigma model, on such a genus
one surface, is independent of the metric of $X$ and so is a topological
invariant in the target space; it is in fact the elliptic genus of $X$.
(See [\landweber] for an introduction to elliptic genera,
and my lecture in that volume
for an explanation of the field theoretic approach to them.)
As the existence of a holomorphic section of $K^{-1/2}$ was
essential here, this construction will not generalize to genus $g>1$.

The twisted models that I will explain below and which will be
the basis for whatever I have to say about mirror manifolds
are the closest analogs of
the usual supersymmetric sigma model for which global fermionic symmetries
exist regardless of the genus.  In particular the closest analog of the
elliptic genus for $g>1$ involves the half-twisted model
introduced at the end of this section.

\subsec{Kahler Manifolds And The Twisted Models}

We now wish to describe the additional
structure -- $N=2$ supersymmetry, to be precise --
that arises if $X$ is a Kahler manifold.
In this case, local complex coordinates on $X$
will be denoted as $\phi^i$; their complex conjugates
are $\phi^{\overline i}=\overline{\phi^i}$.  ($\phi^I$ will still
denote local real coordinates, say the real and imaginary parts of
the $\phi^i$.)
The complexified
tangent bundle $TX$ of $X$ has a decomposition as $TX=T^{1,0}X
\oplus T^{0,1}X$.  The projections of $\psi_+$ in
$K^{1/2}\otimes \Phi^*(T^{1,0}X)$ and $K^{1/2}\otimes \Phi^*(T^{0,1}X)$,
respectively, will be denoted as $\psi_+^i$ and $\psi_+^{\overline i}$.
Likewise the projections of $\psi_-$ in
$\overline K^{1/2}\otimes \Phi^*(T^{1,0}X)$ and $\overline K^{1/2}
\otimes \Phi^*(T^{0,1}X)$
will be denoted as
$\psi_-^i$ and $\psi_-^{\overline i}$, respectively.
The Lagrangian can be written
$$L=2t\int_\Sigma d^2z\left({1\over 2}g_{IJ}\partial_z\phi^I\partial_{\bar z}
\phi^J+i\psi_-^{\overline i}D_z\psi_-^ig_{\overline i i}
+i\psi_+^{\overline i}D_{\overline z}\psi_+^ig_{\overline i i}
+R_{i\,\overline i j\,\overline j}\psi_+^i\psi_+^{\overline i}
\psi_-^j\psi_-^{\overline j}\right). \etag\dalfo$$

As for the fermionic symmetries, these are twice as numerous as before
because of the Kahler structure; this is analogous to the decomposition
of the exterior derivative on a complex manifold
as $d=\overline\partial+\partial$.
I will write the following formulas
out in detail because we will need various
specializations of them in describing the twisted models.
In terms of infinitesimal fermionic parameters $\alpha_-,\widetilde\alpha_-$
(which are holomorphic sections of $K^{-1/2}$)
and $\alpha_+,\widetilde\alpha_+$ (antiholomorphic sections of $\overline
K^{-1/2}$), the transformation laws are
$$\eqalign
{\delta \phi^i & = i\alpha_-\psi_+^i+i\alpha_+\psi_-^i \cr
 \delta \phi^{\overline i} & = i\widetilde\alpha_-\psi_+^{\overline i}
             +i\widetilde\alpha_+\psi_-^{\overline i} \cr
 \delta \psi_+^i & =-\widetilde\alpha_-\partial_z\phi^i-i\alpha_+
     \psi_-^j\Gamma^i_{jm}\psi_+^m \cr
 \delta \psi_+^{\overline i} & =-\alpha_-\partial_z\phi^{\overline i}
-i\widetilde \alpha_+
     \psi_-^{\overline j}\Gamma^{\overline i}_{\overline j\overline m}
\psi_+^{\overline m} \cr
 \delta \psi_-^i & =-\widetilde\alpha_+\partial_{\overline z} \phi^i-i\alpha_-
     \psi_+^j\Gamma^i_{jm}\psi_-^m \cr
 \delta \psi_-^{\overline i} & =
-\alpha_+\partial_{\overline z}\phi^{\overline i}-i\widetilde\alpha_-
     \psi_+^{\overline j}\Gamma^{\overline i}_{\overline j\overline m}
\psi_+^{\overline m} .\cr
}        \etag\elfo$$

Now, the twisted models are constructed as follows:

(1) Instead of taking $\psi_+{}^i$ and $\psi_+{}^{\overline i}$
to be sections of $K^{1/2}\otimes \Phi^*(T^{1,0}X)$ and $K^{1/2}\otimes
\Phi^*(T^{0,1}X)$, respectively, we take them to be sections of
$\Phi^*(T^{1,0}X)$ and $K\otimes \Phi^*(T^{0,1}X)$,
respectively.  I will call this a $+$ twist.  The terms in the
Lagrangian containing $\psi_+$ are unchanged (except that $D_{\bar z}$
must now be interpreted as the $\bar\partial$ operator of the appropriate
bundle).
Alternatively, we can make what I will call  a $-$ twist,
taking $\psi_+^i$ and $\psi_+^{\overline i}$ to be sections
of $K\otimes \Phi^*(T^{1,0}X)$ and $\Phi^*(T^{0,1}X)$, respectively.

(2) Similarly, we twist $\psi_-$ by either a $+$ twist,
taking $\psi_-^i$ to be a section of $\Phi^*(T^{1,0}X)$ and
$\psi_-^{\overline i}$ to be a section of $\bar K\otimes\Phi^*(T^{0,1}X)$,
or a $-$ twist, taking $\psi_-^i$ to be a section of
$\overline K\otimes \Phi^*(T^{1,0}X)$ and
$\psi_-^{\overline i}$ to be a section of $\Phi^*(T^{0,1}X)$.
Again the Lagrangian is unchanged ($D_z$ is now interpreted as
the $\partial$ operator of the appropriate bundle).

If one is twisting only $\psi_+$ or only $\psi_-$, it does not much matter
if one makes a $+$ twist or a $-$ twist.  The two choices
differ by a reversal of the complex structure of $X$, and we are interested
in considering all possible complex structures anyway.
However, when we twist both
$\psi_+$ and $\psi_-$, there are two essentially different theories
that can be constructed.  By making a $+$ twist of $\psi_+$
and a $-$ twist of $\psi_-$, we make what I will call the
$A$ theory.  By making $-$ twists of both $\psi_+$ and $\psi_-$
we make what I will call the $B$ theory.  When I want to make the
dependence on $X$ explicit, I will call these theories $A(X)$ and $B(X)$.

Locally the twisting does nothing at all, since locally
$K$ and $\overline K$ are trivial anyway.  In particular,
in the twisted models, the transformation laws \elfo\ are still valid,
but globally the parameters
 $\alpha,\overline \alpha$, etc., must be interpreted as sections
of different line bundles.  For instance, in the $A$ model,
$\alpha_-$ and $\widetilde\alpha_+$ are functions while
$\alpha_+$ and $\widetilde\alpha_-$ are sections of $\overline K^{-1}$ and
$K^{-1}$.  One can   therefore canonically pick $\alpha_-$
and $\widetilde\alpha_+$ to be constants (and the others to vanish);
this gives canonical global fermionic symmetries  of the $A$ model that are
responsible for its simplicity.
Similarly the $B$ model has two canonical global fermionic symmetries.
These global fermionic symmetries are nilpotent and behave
as BRST-like symmetries.

There is an obvious variant that is possible, and that is to twist only
$\psi_+$ or only $\psi_-$, leaving the other untwisted.  I will call
this the half-twisted model.  The half-twisted model has only one canonical
fermionic symmetry, just the situation which for $g=1$ usually leads
to the elliptic genus.  The half-twisted model would appear
to be the most reasonable framework for generalizing the elliptic genus
to $g>1$.  It also is of phenomenological interest in the following
sense.  In superstring compactifications on Calabi-Yau
manifolds of $\dim_\IC X=3$, the $A$ model is suitable for computing
the $\overline{\bf 27}^3$ Yukawa couplings and the $B$ model for
computing the ${\bf 27}^3$ Yukawa couplings.  To get a full understanding
of the low energy theory, one also needs the ${\bf 1}\cdot {\bf 27}\cdot
\overline {\bf 27}$ and ${\bf 1}^3$ Yukawa couplings; these are not
naturally studied in either the $A$ or $B$ model but involve
BRST invariant observables of
the half-twisted model.  The analysis below of the
$A$ model can be applied to the half-twisted model to show that
correlations of BRST observables reduce to a sum over tree level
computations in instanton fields.

Mirror symmetry (since it reverses one of the $U(1)$ quantum numbers
in the $N=2$ superconformal algebra) can be taken to exchange
the  $+$ and $-$
twists of $\psi_+$ while leaving $\psi_-$ alone.
As a result, mirror symmetry exchanges $A$ and $B$ models (which
differ by the choice of twist of $\psi_+$, for fixed twist of $\psi_-$)
and
maps a half-twisted model to another half twisted model
(since in these models one makes no twist of $\psi_-$ anyway).

Since the elliptic genus of a complex manifold is the same as the
partition function of the half-twisted (or untwisted)
model on a Riemann surface of
genus one, mirror pairs have the same elliptic genus.

We now turn to a detailed description of the $A$ and $B$ models.

\newsec{The $A$ Model}

In the $A$ model, we regard $\psi_+^{i}$ and $\psi_-^{\overline i}$
as sections of $\Phi^*(T^{1,0}X)$ and
$\Phi^*(T^{0,1}X)$, respectively.
It is convenient to combine them into a section $\chi$ of
$\Phi^*(TX)$ (so henceforth $\chi^i=\psi_+^i$, and $\chi^{\overline i}
=\psi_-^{\overline i}$).
As for $\psi_+^{\bar i}$, it is in the $A$ model a $(1,0)$ form on
$\Sigma$ with values in $\Phi^*(T^{0,1}X)$; we will denote it
as $\psi_z^{\bar i}$.  On the other hand, $\psi_-^{i}$ is now
a $(0,1)$ form with values in $\Phi^*(T^{1,0}X)$, and will be denoted
as $\psi_{\overline z}^{i}$.

The topological transformation laws are found from \elfo\ by setting
$\alpha_+=\widetilde\alpha_-=0$ and setting $\alpha_-$ and $\widetilde
\alpha_+$ to constants, which we will call $\alpha$ and $\widetilde\alpha$.
The result is
$$\eqalign{
\delta\phi^i  &   = i\alpha \chi^i \cr
\delta\phi^{\overline i}&=i\widetilde\alpha\chi^{\overline i} \cr
\delta\chi^i&=\delta\chi^{\overline i} = 0 \cr
\delta\psi_z^{\bar i}& = -\alpha\partial_z\phi^{\bar i}-
i\tilde\alpha\chi^{\bar j}\Gamma^{\bar i}_{\bar j\bar m}
         \psi_z^{\bar m} \cr
\delta\psi_{\overline z}^{i} & = -\tilde\alpha\partial_{\overline z}
\phi^{i}-i\alpha\chi^{j}\Gamma^{i}
_{jm}\psi_{\overline z}^m.\cr} \etag\gulfo$$
The supersymmetry
algebra of the original model collapses for these topological
transformation laws to $\delta^2=0$, which holds modulo the equations
of motion.  (By including auxiliary fields, as in [\oldwitten], one can
get $\delta^2=0$ off-shell.)

Henceforth, we will generally for simplicity set $\alpha=\widetilde\alpha$.
(The additional structure that we will overlook is related
to the Hodge decomposition of the cohomology of the moduli space of
holomorphic maps of $\Sigma$ to $X$.)  In this case, the first two
lines of \gulfo\ combine to $\delta \Phi^I=i\alpha\chi^I$.
Also, we will sometimes express
the transformation laws in terms of the BRST operator
$Q$, such that $\delta W=-i\alpha\{Q,W\}$ for any field $W$.  Of course
$Q^2=0$.

In terms of these variables, the Lagrangian is simply
$$L=2t\int_\Sigma d^2z \left({1\over 2}g_{IJ}\partial_z\phi^I\partial_{\bar z}
\phi^J+i\psi_{z}^{\overline i}D_{\bar z}\chi^ig_{\overline i i}
+i\psi_z^{i}D_{z}\chi^{\bar i}g_{\overline i i}
-R_{i\overline i j\overline j}\psi_{\bar z}^i\psi_z^{\overline i}
\chi^j\chi^{\overline j}\right). \etag\ddalfo$$
It is now a key fact that this can be written modulo terms that vanish
by the $\psi$ equation of motion as
$$L=it\int_\Sigma d^2z \,\,\{Q,V\}+t\int_\Sigma \Phi^*(K) \etag\polfo$$
where
$$V=g_{i\overline j}\left(\psi_z^{\overline i}
\partial_{\bar z}\phi^{j}
+\partial_z\phi^{\bar i}\psi_{\overline z}^{j}\right), \etag\tolfo$$
while
$$\int_\Sigma\Phi^*(K)=\int_\Sigma  d^2z\,\,\left(\partial_z\phi^i\partial
_{\overline z}\phi^{\overline j}g_{i\bar j}-\partial_{\overline z}\phi^i
\partial_z\phi^{\overline j}g_{i\overline j}\right) \etag\elfo$$
is the integral of the pullback of the Kahler form $K=-ig_{i\bar j}
dz^idz^{\bar j}$.
Thus $\int\Phi^*(K)$ depends only on the cohomology class of $K$
and the homotopy class of the map
$\Phi$.  If, for instance, $H^2(X,\IZ)\cong \IZ$, and the metric $g$
is normalized so that the periods of $K$ are integer multiples of $2\pi$,
then
$$\int_\Sigma\Phi^*(K)=2\pi n,       \etag\ulfo$$
where $n$ is an integer, the instanton number or degree.  We will
adopt this terminology for simplicity; this involves no essential
distortion.

Instead of saying that \polfo\ is true modulo the $\psi$ equations
of motion, we could modify the BRST transformation law of $\psi$
(by adding terms that vanish on shell) to make \polfo\ hold exactly.
We will not spell out the requisite additional terms in the transformation
law, which do not affect the analysis below, since the operators
${\cal O}_a$ that we will consider
are independent of $\psi$.

\subsec{Reduction To Weak Coupling}

{}From equation \polfo, we can give a quick explanation of one of the
key properties of the model, which is the reduction to weak coupling.
Suppose that we wish to calculate the path integral for fields of
degree $n$.  With insertions of some BRST invariant operators
${\cal O}_a$ (the details of which we will discuss presently),
one wishes to compute
$$\langle\prod_a{\cal O}_a\rangle_n
  =e^{-2\pi n t}\int_{B_n} D\phi\,\,D\chi\,\,D\psi\,\,\,e^{-it\{Q,\int V\}}
\cdot \prod_a{\cal O}_a.     \etag\colfo$$
Here $B_n$ is the component of the field space for maps of degree
$n$, and $\langle ~~\rangle_n$ is the degree $n$ contribution
to the expectation value. We have made use of \ulfo\ to pull out
an explicit factor $e^{-2\pi nt}$ which will turn out to contain the
entire $t$ dependence of $\langle~~~\rangle_n$.

Standard arguments using the $Q$ invariance and the fact that $Q^2=0$
show that $\langle \{Q,W\}\rangle_n=0$ for any $W$.  It is therefore
also true that, as long as $\{Q,{\cal O}_a\}=0$ for all $a$,
\colfo\ is invariant under ${\cal O}_a\to {\cal O}_a+\{Q,S_a\}$ for
any $S_a$.
Thus, the ${\cal O}_a$ should be considered as representatives of
BRST cohomology classes.

Likewise, and this is the key point, \colfo\ is independent of $t$
(as long as ${\rm Re}\,\,\,t>0$ so that the path integral converges)
except for the explicit factor of $e^{-2\pi nt}$ that has been  pulled
out.   In fact, differentiating the other $t$ dependent factor
$\exp(-it\{Q,\int V\})$ with respect to $t$ just brings down irrelevant
factors of the form $\{Q,\dots\}$.  Therefore, the path integral
in \colfo\
can be computed by taking the limit of large ${\rm Re}\,\,t$.
This is the conventional weak coupling limit (for maps of degree $n$).

Looking back at the original form of the Lagrangian \dalfo, or for
that matter at the form of $V$, one sees that for given $n$, the bosonic
part of $L$ is minimized for holomorphic maps of $\Sigma$ to $X$,
that is maps
obeying
$$ \partial_{\bar z}\phi^i=\partial_z\phi^{\bar i}=0. \etag\holo$$
The weak coupling limit therefore involves a reduction to the moduli
space ${\cal M}_n$ of holomorphic maps of degree $n$.
The entire path integral, for maps of degree $n$,
reduces to an integral over ${\cal M}_n$ weighted by one loop determinants
of the non-zero modes.  (A possibly more fundamental
explanation of this reduction will be given in \S5.)
In particular, $\langle \dots \rangle_n$ vanishes for $n<0$, as there
are no holomorphic maps of negative degree.

We can also now explain why the model is a topological field theory,
in the sense that correlation functions $\langle\prod_a{\cal O}_a\rangle$
are independent of the complex structure of $\Sigma$ and $X$,
and depend only on the cohomology class of the Kahler form $K$.
This is certainly true of $\int_\Sigma \Phi^*(K)$.
For the rest, all dependence
of the Lagrangian on the complex structure of $\Sigma$ or $X$ is buried
in the definition of $V$, which appears in the path integral
only in the form $\{Q,V\}$; varying the path integral with respect
to the complex structure of $\Sigma $ or $X$ will therefore bring
only irrelevant factors of the form $\{Q,\dots\}$.

\subsec{The Ghost Number Anomaly}

The Lagrangian \ddalfo\ has at the classical level a ``ghost number''
conservation law, with $\chi$ having ghost number $1$, $\psi$
having ghost number $-1$, and $\phi$ having ghost number 0.
The BRST operator $Q$ has ghost number 1.

At the quantum level, the ghost number is not really a symmetry
because of the anomaly associated with the index or Riemann-Roch theorem.
Let $a_n$ be the number of $\chi$ zero modes, that is the dimension
of the space of solutions of the equations $D_{\bar z}\chi^i
=D_z\chi^{\bar i}=0$.  Similarly, let $b_n$ be the number of
$\psi$ zero modes, solutions of $D_{\bar z}\psi_{z}^{\bar i}
=D_z\psi_{\bar z}^i=0$.\foot{Though not indicated in our notation,
$a_n$ and $b_n$ may depend on the particular map $\Phi$ considered;
we return to this presently.}   The index theorem gives a simple formula
for the difference $w_n=a_n-b_n$.  In particular, $w_n$ is a topological
invariant.  For instance, if $X$ is a Calabi-Yau manifold of complex
dimension $d$, and $\Sigma$ has genus $g$,
then $w_n=2d(1-g)$, independent of $n$.
The expression $\langle \prod{\cal O}_a\rangle_n$ will vanish unless the sum
of the ghost
numbers of the ${\cal O}_a$ is equal to $w_n$.

An essential fact is that the equation for a $\chi$ zero mode
is precisely the linearization of the instanton equation \holo,
and consequently the space of $\chi$ zero modes is precisely
$T{\cal M}_n$, the tangent space to ${\cal M}_n$.  In particular,
if ${\cal M}_n$ is a smooth manifold, then $a_n$ is its (real) dimension
and in particular is a constant.

The number $w_n=a_n-b_n$ is often called the ``virtual dimension'' of
${\cal M}_n$.  The reason for this terminology is that in a sufficiently
generic situation (which may be unattainable in complex geometry) one
would expect that if $w_n>0$ (the only situation that will be of interest)
then $b_n=0$ and hence $w_n=a_n$.

Somewhat more generally, as long as ${\cal M}_n$ is smooth so that
$a_n$ is a constant, $b_n$ will also be constant.
Hence the space $V$ of
$\psi$ zero modes will vary as the fibers of a vector
bundle ${\cal V}$ over ${\cal M}_n$.  At singularities of ${\cal M}_n$,
$a_n$ and $b_n$ may jump.

\subsec{Observables Of The $A$ Model}

To prepare for actual calculations, we need a preliminary discussion
of the observables of the $A$ model.

The BRST cohomology of the $A$ model, in the space of local operators,
can be represented by operators that are functions of $\phi$ and
$\chi$ only.\foot{This happy circumstance prevents
complications related to the fact that \polfo\ only holds on shell
or after modifying the $\psi$ transformation laws.}
They have the following simple construction.

Let $W=W_{I_1I_2\dots I_n}(\phi)d\phi^{I_1}d\phi^{I_2}\dots d\phi^{I_n}$ be an
$n$-form on $X$.  We can define a corresponding local operator
$${\cal O}_W(P)=W_{I_1I_2\dots I_n}\chi^{I_1}\dots \chi^{I_n}(P).\etag\joj$$
The ghost number of ${\cal O}_W$ is $n$.
A simple calculation shows that
$$\{Q,{\cal O}_W\}=-{\cal O}_{dW}, \etag\poj$$
with $d$ the exterior derivative on $W$.  Therefore, taking
$W\to {\cal O}_W$ gives a natural map  from the de Rham cohomology
of $X$ to the BRST cohomology of the quantum field theory $A(X)$.
If one restricts oneself to local operators (a more general class
is considered in \S7), this map is an isomorphism.

Particularly convenient are the following representatives of the
cohomology.  Let $H$ be a submanifold of $X$ (or more generally
any homology cycle).  The ``Poincar\'e dual'' of $H$ is a cohomology
class that counts intersections with $H$.  It can be represented
by a differential form $W(H)$ that has delta function support on $H$.
Hopefully it will cause no confusion if we refer to ${\cal O}_{W(H)}$
as ${\cal O}_H$.
The ghost number of ${\cal O}_H$ is the codimension of $H$.

\subsec{Evaluation Of The Path Integral}

Now let us carry out the evaluation of the path integral.  We pick
some homology cycles $H_a,\,\,\,a=1\dots s$, of codimensions $q_a$.  We also
pick points $P_a\in \Sigma$.  We want
to compute the quantity
$$\langle {\cal O}_{H_1}(P_1)\dots {\cal O}_{H_s}(P_s)\rangle_n
=e^{-2\pi nt}\int_{B_n}D\phi\,\,D\chi\,\,D\psi\,\,
e^{it\int \{Q,V\}}\cdot \prod {\cal O}_{H_a}(P_a).  \etag\roofo$$
This quantity will vanish unless $w_n$, the virtual dimension of
moduli space, is equal to $\sum_aq_a$.

The path integral in \roofo\ reduces, upon using the independence of $t$
and taking ${\rm Re}\,\,t\to \infty$,
to an integral over the moduli space ${\cal M}_n$
of instantons.  Moreover, as we have picked ${\cal O}_{H_a}(P_a)$
to have delta function support for instantons $\Phi$ such that
$$\Phi(P_a)\in H_a,\etag\murfo$$
the path integral
actually reduces to an integral over the moduli space $\tilde{\cal M}_n$
of instantons obeying \murfo.

In a ``generic'' situation, the dimension $a_n$ of ${\cal M}_n$ coincides
with the virtual dimension $w_n$.  Moreover, requiring
$\Phi(P_a)\in H_a$ involves imposing $q_a$ conditions.  Hence
``generically'' the dimension of $\tilde {\cal M}_n$ should be
$w_n-\sum_aq_a=0$.  In such a case, $\tilde{\cal M}_n$ will consist
of a finite set of points.  Let $\#\tilde{\cal M}_n$ be the number
of such points.  In determining the contribution of any of those
points to the path integral, we can take ${\rm Re}\,\,t\to\infty$.
The computation reduces to evaluation of a ratio of boson and fermion
determinants; this ratio is however simply equal to $+1$, because
of the BRST symmetry which ensures cancellation
between bose and fermi modes.\foot{In general
such a BRST symmetry ensures that the ratio of fermion determinants,
in expanding around a BRST fixed point, is $+1$ or $-1$.  In the present
case, the ratio is $+1$ since boson determinants are always positive,
and the fermion determinant of the $A(X)$ model is also positive
since the $\chi^i,
\psi_z^{\bar i}$ determinant is the complex conjugate of the $\chi^{\bar i}$,
$\psi_{\bar z}^i$ determinant. }

In a generic
situation, we therefore have
$$\langle \prod_{a=1}^s{\cal O}_{H_a}(P_a)\rangle_n=e^{-2\pi nt}\cdot
\#\tilde{\cal M}_n.         \etag\golfox$$
Summing over $n$ we get ``generically''
$$\langle \prod_a^s{\cal O}_{H_a}(P_a)\rangle=\sum_{n=0}^\infty
e^{-2\pi nt}\cdot
\#\tilde{\cal M}_n.         \etag\golfox$$

In complex geometry, life is not always ``generic'' and $\widetilde{\cal M}_n$
may well have components of positive dimension.  Suppose $\widetilde{\cal M}_n$
has real
dimension $s$ (or focus on a component of that dimension).
If so, by virtue of the Riemann-Roch theorem, the space $V$ of $\psi$ zero
modes
is $s$ dimensional and varies as the fibers of a vector bundle
${\cal V}$ of (real) dimension $s$ over $\widetilde{\cal M}_n$.
\foot{The $\psi$ zero modes were discussed in the subsection on
the ghost number anomaly; however,
the equation for such zero modes must now be corrected to permit
poles at $P_a$ tangent to $H_a$.  A general extension of index theory
to such situations, with the properties essential here,
has been given by Gromov and Shubin [\gromov].}
It can be argued on rather general grounds that the generalization
of counting the number of points in $\tilde{\cal M}_n$ is the evaluation
of the Euler class $\chi({\cal V})$ of the bundle ${\cal V}$:
$$\#\tilde{{\cal M}}_n\to \int_{\tilde{{\cal M}}_n}\chi({\cal V}).
    \etag\bopo$$
I refer to \S3.3 of [\witten]
for an explanation
of this key point (and a detailed field theoretic calculation showing
explicitly how $\chi({\cal V})$ arises in a representative
example); also, see
[\atiyah] for some of the background.

The generalization of \golfox\ is therefore
$$\langle \prod_{a=1}^s{\cal O}_{H_a}(P_a)\rangle=\sum_{n=0}^\infty
e^{-2\pi nt}\cdot \int_{\tilde{\cal M}_n}
\chi({\cal V}).  \etag\ggolfox$$
In a real life situation, involving multiple covers
of an isolated rational curve in a three dimensional Calabi-Yau
manifold, the Euler class of ${\cal V}$ has been evaluated by
Aspinwall and Morrison [\aspinwall], who as a result were able to justify
a formula that had been guessed empirically by Candelas et. al. [\candelas].

Notice that in deriving \golfox\ and \ggolfox\ we have not assumed
that $\Sigma$ has genus zero.  The restriction to genus zero will
arise only when (in \S6) we explain the relation of these correlation
functions of the twisted model to ``physical'' correlation functions
of the untwisted model.

Although it may be impossible in complex geometry
to achieve a ``generic'' situation
(in which the actual dimension of ${\cal M}$ coincides with its virtual
dimension),
this can always be done by perturbing the
complex structure of $X$ to a generic non-integrable almost complex
structure.  (This is allowed in the $A$ model [\oldwitten].)
The importance of
\ggolfox\ comes from the fact that it is generally impractical to do
calculations based on generic non-integrable deformations.
The generic non-integrable deformations are sometimes useful
for theoretical arguments; see the end of \S7.
See [\gromov] for the theory of almost holomorphic curves in
almost complex manifolds.

\newsec{The $B$ Model}

Now we will consider the $B$ model in a similar spirit.

In the $B$ model, $\psi_{\pm}^{\bar i}$ are sections of $\Phi^*(T^{0,1}X)$,
while $\psi_+^i$ is a section of $K\otimes \Phi^*(T^{1,0}X)$, and
$\psi_-^i$ is a section of $\overline K\otimes \Phi^*(T^{1,0}X)$.
It is convenient to set
$$\eqalign{\eta^{\bar i} & = \psi_+^{\bar i}+\psi_-^{\bar i} \cr
           \theta_i & = g_{i\bar i}\left(\psi_+^{\bar i}-\psi_-^{\bar i}\right)
               .    \cr}         \etag\upo$$
Also, we combine $\psi_{\pm}^i$ into a one form $\rho$ with values
in $\Phi^*(T^{1,0}X)$; thus, the $(1,0)$ part of $\rho$ is
$\rho_z^i=\psi_+^i$, and the $(0,1)$ part of $\rho$ is $\rho_{\bar z}^i
=\psi_-^i$.

As for the supersymmetry transformations, we now set $\alpha_\pm=0$,
and set $\tilde\alpha_\pm$ to constants; in fact, for simplicity we will
just set $\tilde\alpha_+=\tilde\alpha_-=\alpha$.
The transformation laws are then
$$\eqalign{\delta \phi^i    &   =      0 \cr
           \delta \phi^{\overline i} & =i\alpha\eta^{\bar i} \cr
            \delta \eta^{\bar i} & =\delta \theta_i = 0 \cr
            \delta \rho^i & = -\alpha \,\,d\phi^i . \cr} \etag\newtr$$
The BRST operator is again defined by $\delta(\dots)=-i\alpha\{Q,\dots\}$,
and obeys $Q^2=0$ modulo the equations of motion.

The Lagrangian is
$$\eqalign{
L=&t\int_\Sigma d^2z\left(g_{IJ}\partial_z\phi^I\partial_{\bar z}\phi^J
+i\eta^{\bar i}(D_z\rho_{\bar z}^i+D_{\bar z}\rho_z^i  )g_{i\bar i}
\right.\cr &\left.
\qquad +i\theta_i(D_{\bar z}\rho_z{}^i-D_z\rho_{\bar z}{}^i)
+R_{i\bar ij\bar j}\rho_z^i\rho_{\bar z}^j\eta^{\bar i}\theta_k g^{k\bar j}
\right).\cr} \etag\xolfo$$
This can be rewritten
$$L=it\int\{Q,V\}+tW        \etag\horro$$
where
$$V=g_{i\bar j}\left(\rho_z^i\partial_{\bar z}\phi^{\bar j}
        +\rho_{\bar z}^i\partial_z\phi^{\bar j}\right)\etag\orro$$
and
$$ W=\int_\Sigma \left(-\theta_i D\rho^i
-{i\over 2}R_{i\bar ij\bar j}\rho^i\wedge \rho^j\eta^{\bar i}\theta_k g^{k\bar
j
   }
\right). \etag\rro$$
Here $D$ is the exterior derivative on $\Sigma$ (extended to act on
forms with values in $\Phi^*(T^{1,0}X)$ by using the pullback of the
Levi-Civita
connection of $X$), and $\wedge$ is the wedge product of forms.

We can now see that the $B$ theory is a topological field theory, in the sense
that it is independent of the complex structure of $\Sigma$
and the Kahler metric of $X$.\foot{The theory definitely depends on
the complex structure of $X$, which enters in the BRST transformation laws.}
Under a change of complex structure of $\Sigma$ or Kahler metric
of $X$, the Lagrangian only changes by irrelevant terms of the form
$\{Q,\dots\}$.  This is obviously true for
the $\{Q,V\}$ term on the right hand side of \horro.
As for $W$, it is entirely independent of the complex structure of $\Sigma$,
since it is written in terms of differential forms.  It is less obvious,
but true, that under change of Kahler metric of $X$, $W$ changes by $\{Q,\dots
\}$.
These observations are ``mirror'' to our
earlier result
that the $A$ theory is independent of the complex structure
of $\Sigma $ and $X$ but depends on the Kahler class of the metric of $X$.

Similarly, the $B$ theory is independent of the coupling constant $t$
(except for a trivial factor which will appear shortly) as long as ${\rm
Re}\,\,
t>0$ so that the path integral converges.  Under a change of $t$,
the $t\{Q,V\}$ term changes by $\{Q,\dots\}$.
As for the $t$ in $tW$, this can be removed by redefining
$\theta\to \theta/t$ (since $V$ is independent of $\theta$ and $W$ is
homogeneous of degree one).  Hence, the theory is independent of $t$ except
for factors that come from the $\theta$ dependence of the observables.
If ${\cal O}_a$ are BRST invariant operators that are homogeneous
in $t$ of degree $k_a$, then the $t$ dependence
of $\langle \prod_a{\cal O}_a\rangle$
is  a factor of $t^{-\sum_ak_a}$, which arises from the rescaling
of $\theta$ to remove the $t$ from $tW$.  This trivial $t$ dependence
of the $B$ theory should be constrasted
with the complicated $t$ dependence of the $A$ theory, coming from
the instanton sum.

Because the $t$ dependence of the $B$ theory is trivial and known,
all calculations can be performed in the limit of large ${\rm Re}\,\,t$,
that is, in the ordinary weak coupling limit.  In this limit, one expands
around minima of the bosonic part of the Lagrangian; these are just the
constant
maps $\Phi:\Sigma\to X$.  The space of such constant maps is a copy of
$X$, so the path integral reduces to an integral over $X$.  We will make
this more
explicit after identifying the observables.

This is to be contrasted with the $A$ theory, in which one has to integrate
over moduli spaces of holomorphic curves.  The difference arises because
in the $A$ theory, the $t$ dependence becomes standard only after
removing a factor of $t\int \Phi^*(K)$, and after doing this, the
rest of the bosonic part of the action is zero for arbitrary holomorphic
curves, not just constant maps.
In \S5, I will give an alternative and perhaps more fundamental explanation
of why calculations in the $B$ theory reduce to integrals over $X$
while in the $A$ theory they reduce to integrals over instanton
moduli space.

\subsec{Anomalies}

The fermion determinant of the $A$ model is real and positive
(as the $\chi^i,\psi_{z}^{\bar i}$ determinant is the complex conjugate
of the $\chi^{\bar i},\psi_{\bar z}^i$ determinant).  In particular,
there is no problem in defining this determinant as a function, and the
$A$ model, even before taking BRST cohomology, makes at least
some sense as a quantum
field theory (perhaps with a cutoff, and not conformally invariant)
for any complex manifold $X$, not necessarily Calabi-Yau.
(In fact, in [\oldwitten], the $A$ model was defined for general
almost complex manifolds.)
The fact that the eventual recipe \ggolfox\ for computing correlation
functions does not use the Calabi-Yau condition is related to this.

The $B$ model is very different.
Because the zero forms $\eta^{\bar i},\,\,\,g^{\bar i i}\theta_i$
are sections of $T^{0,1}X$ and the
one forms $\rho^i$ are sections of $T^{1,0}X$, the fermion determinant
in the $B$ model is complex.  The $B$ model does not make any sense
as a quantum field theory, even with cutoff, without an anomaly cancellation
condition that makes it possible to define the fermion determinants
as functions (not just sections of some line bundle).  The relevant
condition is $c_1(X)=0$, that is, $X$ should be a Calabi-Yau
manifold.\foot{In two dimensions, anomalies are quadratic
in the coupling
of fermions to gravitational and gauge fields.  To get an anomaly
that depends on the twisting (since the untwisted model is not anomalous)
and on $X$ (since the twisted model is a non-anomalous free field theory
for $X=\IC^n$), we must consider a term linear
in the gravitational field, that is
the spin connection of $\Sigma$, and linear in the gauge field,
that is the pull-back of the Levi-Civita connecton  of $X$.  The
only invariant linear in the latter is $c_1(X)$, and standard considerations
show that $c_1(X)$ is indeed the obstruction to defining the fermion
determinant
in the $B$ model.}
Thus, in the $B$ model, the Calabi-Yau condition plays an even more
fundamental role than it does in the untwisted model, where it is merely
necessary for conformal invariance.

Like the $A$ model, the $B$ model has an important $\IZ$ grading by
a quantum number that we will call the ghost number.  The ghost
number is $1$ for $\eta$ and $\theta$, $-1$ for $\rho$, and zero
for $\phi$.  $Q$ is of degree 1.  If $X$ is a Calabi-Yau manifold
of complex dimension $d$, and ${\cal O}_a$ are BRST invariant operators
of ghost number $w_a$, then $\langle{\cal O}_a\rangle$ vanishes
in genus $g$ unless
$$\sum_aw_a=2d(1-g).         \etag\gippo$$
(There is actually a more refined $\IZ\times \IZ$ grading, which
we have obscured by setting $\tilde\alpha_+=\tilde\alpha_-$
and combining $\psi_\pm^i$ into $\rho$.)

\subsec{The Observables}

Now we wish to make the simplest observations about the observables
of the $B$ model, analogous to our earlier discussion of the $A$ model.

Instead of the cohomology of $X$, as in the $A$ model, we consider
$(0,p)$ forms on $X$ with values in $\wedge^qT^{1,0}X$, the $q^{th}$
exterior power of the holomorphic tangent bundle of $X$.\foot{In complex
geometry, $T^{1,0}X$ might be called simply $TX$, but we have used that name
for the complexification of the real tangent bundle of $X$.}  Such an object
can be written
$$V=d\bar z^{i_1}d\bar z^{i_2}\dots d\bar z^{i_p}V_{\bar i_1\,\bar i_2
\dots \bar i_p}{}^{j_1j_2\dots j_q}{\partial\over\partial z_{j_1}}
\dots {\partial\over\partial z_{j_q}}           \etag\kolpo$$
($V$ is antisymmetric in the $j$'s as well as in the $\bar i$'s.)
The sheaf cohomology group $H^p(X,\wedge^qT^{1,0}X)$ consists
of solutions of $\bar\partial V=0$ modulo $V\to V+\bar \partial S$.

For every $V$ as in \kolpo, and $P\in \Sigma$, we can form the
quantum field theory operator
$${\cal O}_V=\eta^{\bar i_1}\dots \eta^{\bar i_p}V_{\bar i_1\dots
\bar i_p}{}^{j_1\dots j_q}\psi_{j_1}\dots \psi_{j_q}.  \etag\olpo$$
One finds that
$$\{Q,{\cal O}_V\}  = -{\cal O}_{\bar\partial V}, \etag\solpo$$
and consequently ${\cal O}_V$ is BRST invariant if $\bar\partial V=0$
and BRST exact if $V=\bar\partial S$ for some $S$.  Thus
$V\to {\cal O}_V$ gives a natural map from $\oplus_{p,q}H^p(X,\wedge^q
T^{1,0}X)$ to the BRST cohomology of the $B$ model.
This is in fact an isomorphism (as long as one considers only
local operators; in \S7, we will make
a slight generalization).

\subsec{Correlation Functions}

Now picking points $P_a\in \Sigma$ and
classes $V_a$ in $H^{p_a}(X,\wedge^{q_a}T^{1,0}X)$,
we wish to compute
$$\langle \prod_a{\cal O}_{V_a}(P_a)\rangle .\etag\rorp$$
We will consider only the case of genus zero.
It will be clear that \rorp \ vanishes unless
$$\sum_a p_a=\sum_aq_a = d .       \etag\hollof$$
(This is related to a more precise grading of the theory, by left-
and right-moving ghost numbers, that was alluded to following
equation \gippo.)

Taking the large $t$ limit, the calculation reduces as explained
earlier to an integral over the space of constant maps $\Phi:
\Sigma\to X$. In addition to the bose zero modes -- the displacements
of the constant map $\Phi$ -- there are fermi zero modes, which
are the constant modes of $\eta$ and $\theta$.
The nonzero bose and fermi modes enter only via their one loop determinants;
these determinants
are independent of the particular constant map $\Phi:\Sigma\to X$ about
which one is expanding,
and so just go into the definition of the string coupling constant.
So one reduces to a computation involving the zero modes only; and
correlation functions of the $B$ model will reduce
to classical expressions.

Once we restrict to the space of zero modes, a function of $\phi$,
$\eta$ and $\theta$ which is of $p^{th}$ order in $\eta$ and $q^{th}$
order in $\theta$ can be interpreted as a $(0,p)$ form on $X$ with
values in $\wedge^qT^{1,0}X$.  This of course is where the ${\cal O}$'s came
from originally.  In multiplying such functions, one automatically
antisymmetrizes on the appropriate indices because of fermi statistics.
Thus $\prod_{a}{\cal O}_{V_a}$ can be interpreted, using \hollof,
as a $d$ form with values in $\wedge^dT^{1,0}X$.
The map
$$\otimes_aH^{p_a}(X,\wedge^{q_a}T^{1,0}X)\to H^d(X,\wedge^dT^{1,0}X)
           \etag\liliop$$
is the classical wedge product.

What remains is to integrate over $X$ the element of $H^d(X,\wedge^dT^{1,0}X)$
obtained this way.
The Calabi-Yau condition is here essential; it ensures that
$H^d(X,\wedge^dT^{1,0}X)$ is non-zero
and  one dimensional.  The space of linear forms on
this space is thus likewise
one dimensional; any such non-zero form
gives a method of ``integration,'' unique up to a constant multiple.
Of course, the path integral of the $B$ model gives formally a method of
evaluating \rorp\ and hence of integrating an element of $H^d(X,
\wedge^dT^{1,0})$;
this procedure formally is unique up to a multiplicative constant
(a correction to the string coupling constant).  We noted in our discussion
of anomalies that the $B$ model is anomalous except for Calabi-Yau manifolds.

The restriction to Calabi-Yau manifolds amounts to the fact that what
can be integrated naturally are top forms or elements of $H^d(X,\Omega^dX)$.
($\Omega^dX$
is the sheaf of forms of type $(d,0)$.)
In general the relation of $\Omega^dX$ and $\wedge^dT^{1,0}X$ is that
they are {\it inverses}, but in the Calabi-Yau case they are both
trivial, and hence isomorphic.  Indeed, multiplication by the square
of a holomorphic $d$ form gives a map from $\wedge^dT^{1,0}X$ to
$\Omega^dX$.  Empirically,
the choice of a holomorphic $d$ form corresponds to the
choice of the string coupling constant, though this relation is still
somewhat mysterious.

\newsec{The Fixed Point Theorem}

In the last section, we explained why calculations in the $A$ model
reduce to integrals over moduli spaces of holomorphic curves, while
calculations in the $B$ model reduce to integrals over spaces of
constant maps (and ultimately to classical expressions).
I will now (as in [\witten], \S3.1) explain
this in an alternative and perhaps more fundamental
way, as a sort of fixed point theorem.

Consider an arbitrary quantum field theory, with some function space
${\cal E}$ over which one wishes to integrate.  Let $F$ be a group
of symmetries of the theory.  Suppose $F$ acts freely on ${\cal E}$.
Then one has a fibration ${\cal E}\to {\cal E}/F$, and by integrating
first over the fibers of this fibration, one can reduce the integral
over ${\cal E}$ to an integral over ${\cal E}/F$.  Provided one considers
only $F$ invariant observables ${\cal O}$, the integration over the fibers is
particularly simple and just gives a factor of ${\rm vol}(F)$ (the volume of
the group $F$):
$$\int_{\cal E}e^{-L}{\cal O}={\rm vol}(F)\cdot \int_{{\cal E}/F}e^{-L}{\cal O}
.\etag\mornob$$

We want to apply this to the case in which $F$ is the $(0|1)$ dimensional
supergroup generated by the BRST operator $Q$.  This case has some very
special features.  The volume of the group $F$ is zero, since for a fermionic
variable $\theta$,
$$\int d\theta \cdot 1 = 0 . \etag\ornob$$
Hence \mornob\ tells us that if $Q$ acts freely,
the expectation value of any $Q$ invariant operator vanishes.

To express this in another way, if $F$ acts freely,
then one can introduce
a collective coordinate $\theta$ for the BRST symmetry.  BRST invariance
tells us that $L$ and ${\cal O}$ are both independent of $\theta$, and since
the $\theta$ integral of a $\theta$-independent function vanishes, the
path integral would vanish.

%\FIG\oone{The space ${\cal E}$ of all fields, the locus ${\cal E}_0$ of
%fixed points of the BRST symmetry, and a tubular neighborhood of the latter.}
In general, $F$ does not act freely, but has a fixed point locus ${\cal E}_0$.
If so, let ${\cal C}$ be an $F$-invariant neighborhood of ${\cal E}_0$
and ${\cal E}'$ its complement.  Then the path integral restricted
to ${\cal E}'$ vanishes, by the above reasoning. So the entire
contribution to the path integral comes from the integral over
${\cal C}$.  Here ${\cal C}$ can be an arbitrarily small neighborhood,
so the result is really a localization formula expressing the path integral
as an integral on ${\cal E}_0$.
The details depend on the structure of $Q$ near ${\cal E}_0$.
If the vanishing of $Q$ near ${\cal E}_0$
is a generic, simple zero, then the fixed point contribution is simply
an integral over ${\cal E}_0$ weighted by the one loop determinants
of the transverse degrees of freedom.  This is analogous to, say,
the Atiyah-Bott fixed point theorem in topology.

Now let us carry this out in the $A$ and $B$ models.
In the $A$ model, the relevant BRST transformation laws read
$$\eqalign{\delta\psi_z^{\bar i} & = -\alpha\partial_z\phi^{\bar i}-
i\tilde\alpha\chi^{\bar j}\Gamma^{\bar i}_{\bar j\bar m}\psi_z^{\bar m} \cr
\delta\psi_{\bar z}^i & = -\tilde\alpha\partial_{\bar z}\phi^i
-i\alpha\chi^j\Gamma^i_{jm}\psi_{\bar z}^m \cr \delta \phi^I &=i\alpha
\chi^I.         \cr} \etag\yopo$$
Requiring $\delta\phi^I=0$,
we get that $\chi^I=0$ for a BRST fixed point, and
setting $\delta\psi=0$, we see that
in addition, a fixed point must have
$$\partial_{\bar z}\phi^i=\partial_z\phi^{\bar i} = 0 .\etag\rff$$
This is the equation for a holomorphic curve, and shows the localization
of the $A$ model on the space of such curves.

The important part of the BRST transformation law of the $B$ model for our
present purposes is
$$\delta \rho^i= -\alpha \,d\phi^i.   \etag\ddd$$
Setting $\delta\rho^i=0$, we see that
the condition for a fixed point is $d\phi^i=0$; that is,
$\Phi:\Sigma\to X$ must be a constant map.  Thus we recover the localization
of the $B$ model on classical, constant configurations.

\newsec{Relation To The ``Physical'' Model}

So far, we have concentrated exclusively on analyzing the twisted
$A$ and $B$ theories and their correlation functions.
Mirror symmetry is however usually applied to the
correlation functions of the untwisted, physical nonlinear sigma
models.  The purpose
of the present section is to explain why certain correlation functions of
the twisted models (either $A$ or $B$; they can be treated together)
are equivalent to certain correlation functions
of the physical models.

In constructing the twisted models from the physical sigma model, we
``twisted'' various fields by $K^{1/2}$ and $\overline K^{1/2}$ ($K$
being the canonical bundle of a Riemann surface $\Sigma$).
To state the relation between the physical and twisted models in one
sentence, it is simply that the models coincide (with a suitable identification
of the observables) whenever $K$ is trivial and we choose $K^{1/2}$
and $\overline K^{1/2}$ to be trivial, since in that case
the twisting did nothing.
Although there are other examples of Riemann surfaces with trivial
canonical bundle that might be considered, the important example
(for standard applications of mirror symmetry) is the
case that $\Sigma$ is a Riemann surface of genus zero with two points
deleted.

Such a surface, of course, can be thought of as a cylinder
with a complete, flat metric, say $ds^2=d\tau^2+d\sigma^2$,
$-\infty<\tau<\infty,\,\,\,\,\,\,0\leq\sigma\leq 2\pi$.
In computing path integrals on such a surface, we must pick initial
and final quantum states, say $|w\rangle$ and $|w'\rangle$.
Considering first the twisted model,
we assume that these are $Q$ invariant, and so are representatives
of suitable BRST cohomology classes.
Picking also points $P_a\in\Sigma,\,\,a=1\dots s$,
and BRST invariant operators ${\cal O}_a$,
we consider the objects
$$\langle w'|\prod_{a=1}^s{\cal O}_a(P_a)|w\rangle         \etag\ildoo$$
which can be represented by path integrals if we wish.

Of course, \ildoo\ can be
interpreted more symmetrically by compactifying $\Sigma$ --
adding points $P$ and $P'$ and conformally rescaling
the metric to bring them to a finite distance.  Then the states
$|w\rangle$ and $|w'\rangle$ will correspond to BRST invariant operators
${\cal O}_w(P)$ and ${\cal O}_{w'}(P')$.
(In the $A$ or $B$ model,
the BRST cohomology is spanned by operators ${\cal O}_V$,
where $V$ is a de Rham
cohomology class or an element of some $H^p(X,\wedge^qT^{1,0}X)$,
respectively; so ${\cal O}_w$ and ${\cal O}_{w'}$ will automatically
be operators of this type for some $V$'s.)
The
matrix element \ildoo\ is then equivalent to
a correlation function
$$\langle {\cal O}_{w'}(P'){\cal O}_{w}(P)
\prod_{a=1}^s{\cal O}_a(P_a)\rangle \etag\wildoo$$
on the compactified surface $\widehat \Sigma$.
This can be evaluated according to the recipes of sections three and four,
for the $A$ or $B$ model as the case may be.

Now we go back to the open surface $\Sigma$, with its trivial
canonical bundle and flat metric, and make the following key observation.
As $K$ is trivial, we can pick $K^{1/2}$ to be trivial.  If we do so,
the twisting by $K^{1/2}$ does nothing.  Hence,
\ildoo\ is equivalent to some matrix element in the untwisted model.
Of course, the untwisted model has a lot of matrix elements
(since the physical states have the multiplicity of a Fock space);
we will get only a few of them this way.
The operators ${\cal O}_a$ will correspond to some bosonic vertex
operators of the untwisted model; in fact, as discussed in general
terms in [\lerche],
if the ${\cal O}_a$ are chosen as harmonic representatives of the
appropriate BRST cohomology classes, they are
standard vertex operators of massless bosons.
(In the language of [\lerche,\vafa], these particular bosonic vertex
operators generate the chiral ring -- in fact, the $ca$ or
$cc$ chiral ring in the case of the $A$ or $B$ model.)
Let us call the harmonic representatives $B_a$.

What about the initial and final states in \ildoo?
Equivalence of the twisted and untwisted models depends on choosing
$K^{1/2}$ (and $\overline K^{1/2}$)
to be trivial.  This means in the language of the
untwisted model that we are working in the Ramond sector; and thus
the initial and final states are fermions (and in fact, harmonic
representatives of the cohomology classes in question would be ground
state fermions of the untwisted model).  Let us call these fermi
states $|f\rangle$ and $|f'\rangle$.  From the point of view of the
untwisted model,
\ildoo\ might be written
$$\langle f'|\prod_{a=1}^sB_a|f\rangle.       \etag\ldoo$$
I stress, though, that in going from the twisted to the untwisted
model on the flat cylinder $\Sigma$, all that we have changed is
the notation.

Now, \ldoo\ is a coupling of two ground state fermions $|f\rangle$ and
$|f'\rangle$ to an arbitrary number of ground state bosons
$B_a$ (which are all from the same chiral ring).
Of the diversity of possible observables of the untwisted model,
these are of particular importance as they determine the ``superpotential.''
\foot{Actually, the cubic terms in the superpotential come from the
case $s=1$ of the above.  To compute higher terms in the superpotential,
one must consider the integrated, two form version of the ${\cal O}$'s,
which we will introduce in the next section.  The analysis of the relation
between twisted and untwisted models on the cylinder is unchanged.}
We have explained how these particular matrix elements can be identified
with observables of the twisted model.

Of course, if we wish we can compactify $\Sigma$ in the context
of the untwisted model, adding points $P$ and $P'$
at infinity and
conformally scaling the metric to bring them to a finite distance.
At this stage the difference between the twisted and untwisted model will
come in, as the isomorphism between them depends on a trivialization
of $K$.  In the untwisted model, when one projects the points at infinity
to a finite distance, the states $|f\rangle,\,\,|f'\rangle$ will be
replaced by fermion vertex operators $V_f,\,V_{f'}$.  Hence,
\ldoo\ has the alternative interpretation as a correlation function
$$\langle V_f(P)V_{f'}(P')\prod_{a=1}^sB_a\rangle
 \etag\doo$$
on the closed surface $\widehat\Sigma$.
Note that in contrast to \wildoo, which arose from the analogous
compactification in the twisted model, here the ``new'' vertex
operators are of a different type from the old ones.  This is possible
because the untwisted model has vastly more observables
than the twisted models.  The fact that the Yukawa couplings are
derived from a cubic form (with symmetry between the bose and fermi
lines), which is usually regarded as a consequence of space-time supersymmetry,
is manifest in the representation
\wildoo\ of the twisted model,
because the ``fermions'' and ``bosons'' are represented
by the same kind of vertex operators.

\newsec{Closer Look At The Observables}

In this section we will, finally, take a closer look at the observables
of the $A$ and $B$ theories.  We will describe a structure -- a hierarchy
of $q$ form obervables for $q=0,1,2$ -- which must exist on general
grounds.  We will then analyze this hierarchy in some detail in the
$A$ and $B$ models.  In the $A$ model we will obtain a simple answer
which moreover has a simple and standard topological description.
The analogous calculation in the $B$ model turns out to be far more
complicated.
We will not push it through to the end, but we will go far enough
to identify the relevant structure, which turns out to be somewhat
novel.

In either the $A$ or $B$ model, we described a family of observables,
say ${\cal O}_V(P)$, where $V$ is a de Rham cohomology class or
an element of some $H^p(X,\wedge^qT^{1,0}X)$, in the $A$ or $B$ model,
and $P$ is a point in a Riemann surface $\Sigma$.  Correlation functions
$$\langle \prod_{a=1}^s{\cal O}_{V_a}(P_a)\rangle \etag\ddo$$
are independent of the $P_a$, because of the topological invariance of the
theory.  We will systematically exploit the consequences of this fact.
In doing so, we want to think of ${\cal O}_V$ as an operator-valued
zero form; to emphasize this we write it as ${\cal O}^{(0)}$.  We fix
a particular $V$ and do not always indicate it in the notation.

Topological invariance of the theory
-- the fact that correlation functions of ${\cal O}^{(0)}(P)$
are independent of $P$ -- means that ${\cal O}^{(0)}$ must be
a closed zero form up to BRST
commutators,
$$d{\cal O}^{(0)}=\{Q,{\cal O}^{(1)}\}, \etag\hopo$$
for some ${\cal O}^{(1)}$.  This formula, read from right to left,
means that the operator-valued one form
${\cal O}^{(1)}$ is BRST invariant up to an exact form.  Hence, we get
new observables in the theory.
If $C$ is a circle in $\Sigma$ (or more generally a one dimensional
homology cycle), then
$$U(C)=\oint_C{\cal O}^{(1)}  \etag\zopo$$
is a BRST invariant observable.

We can repeat this procedure.  Topological invariance means that correlation
functions of $U(C)$ must be invariant under small displacements of
$C$; this means that ${\cal O}^{(1)}$ must be a closed form up
to BRST commutators,
$$ d{\cal O}^{(1)}=\{Q,{\cal O}^{(2)}\}, \etag\xopo$$
for some ${\cal O}^{(2)}$. Also, \xopo\ means that the two form
${\cal O}^{(2)}$ is BRST invariant up to an exact form, so
$$W=\int_\Sigma {\cal O}^{(2)} \etag\lopo$$
is a new BRST invariant observable.

In this procedure, if ${\cal O}^{(0)}$ has ghost number $q$,
then ${\cal O}^{(i)}$ has ghost number $q-i$, for $i=1,2$.

Obviously, if $X,Y$ are a mirror pair of Calabi-Yau manifolds,
then the mirror symmetry $A(X)\cong B(Y)$ can be applied to the
new observables that we have just described.
This is likely to be particularly interesting for applications of mirror
symmetry with target spaces of complex dimension greater than three.

Now, \xopo\ actually leads to the existence of a more general family
of topological quantum field theories.  If $L$ is the original
Lagrangian, and ${\cal O}_{V_a}^{(0)}$ are the operator valued
zero forms, of ghost number $q_a$,  with which the above procedure begins, then
we get a family of topological Lagrangians,
$$L\to L+\sum_at_a\int_\Sigma {\cal O}_{V_a}^{(2)}. \etag\nwfamily$$
Let us call this the topological family.
As the ghost number of ${\cal O}_{V_a}^{(2)}$ is $q_a-2$, Lagrangians
in the topological family do not necessarily conserve ghost number
(even at the classical level); those that do not are not twistings
of standard renormalizable sigma models.  In the case of a mirror pair,
the whole topological family $A(X)$ is equivalent to the topological
family $B(Y)$.  So far mirror symmetry has been applied only to the subfamilies
of theories that conserve ghost number classically.  It is very likely
that aspects of mirror symmetry that are now not well understood
-- like the nature of the mirror map between the moduli spaces -- are
more transparent in the context of the full topological family.

\subsec{The $A$ Model}

Now we will work out the details of the above for the $A$ model.
This is easy enough.
If
$${\cal O}^{(0)}=V_{I_1I_2\dots I_n}\chi^{I_1}\chi^{I_2}\dots \chi^{I_n},
       \etag\dof$$
for some $n$ form $V$, then $d{\cal O}^{(0)}=\{Q,{\cal O}^{(1)}\}$,
where
$${\cal O}^{(1)} = -nV_{I_1I_2\dots I_n}d\phi^{I_1}\chi^{I_2}\dots
\chi^{I_n}. \etag\hillo$$
And $d{\cal O}^{(1)}=\{Q,{\cal O}^{(2)}\}$, with
$${\cal O}^{(2)}= -{n(n-1)\over 2}V_{I_1I_2\dots I_n}d\phi^{I_1}\wedge
d\phi^{I_2}\chi^{I_3}\dots \chi^{I_n}.              \etag\olof$$
%%%%%%%%%%%%

The above field theoretic formulas correspond to the following topological
construction.  Let ${\cal M}$ be the moduli space of holomorphic
maps of $\Sigma$ to $ X$ of some given homotopy type.  Thus
we have a family of maps $\Phi:\Sigma\to X$ parameterized by ${\cal M}$.
Alternatively, one can think of this as a single map $\Phi:\Sigma\times
{\cal M}\to X$.  Given now an
$n$ dimensional cohomology class $V\in H^*(X)$, we can pull it back
to $\Phi^*(V)\in H^*(\Sigma\times{\cal M})$.  This is an $n$ dimensional
cohomology class of $\Sigma\times {\cal M}$.  To get cohomology classes
of ${\cal M}$, let $\gamma$ be an $s$ dimensional submanifold of
$\Sigma$ (for $s=0,1,$ or 2) and let $i:\gamma\to\Sigma$ be the inclusion.
Then by integration over $\gamma$, one gets
an $n-s$ dimensional class in the cohomology of ${\cal M}$, namely
$$i_*(\Phi^*(V))=\int_\gamma\Phi^*(V).  \etag\pofo$$
For instance, if $\gamma$ is a point $P\in \Sigma$, then integration over
$P$ just means restricting to $P$, and \pofo\ corresponds in the quantum
field theory description to our old friend ${\cal O}_V^{(0)}(P)$.
For $\gamma$ a one-cycle (say a circle $C$) or a two-cycle (which
must be a multiple of $\Sigma$ itself), we get the topological
counterparts of the objects introduced in
\zopo\ and \lopo\ above.

It is not to hard to verify the precise
correspondence between the field theoretic and topological definitions.
See [\baulieu] for further discussion of some of these matters.

\subsec{The $B$ Model}

To understand the analogous issues in the $B$ model are more difficult.
In fact, because the computations involved are rather elaborate,
I will first make some qualitative remarks to indicate what
must be expected.  Then we will just make a few illustrative
computations which indicate the form of the general answer.

The operator-valued zero forms ${\cal O}^{(0)}$ of the $B$ model
are determined by elements of $H^p(X,\wedge^qT^{1,0})$ for various $p$ and
$q$.  The corresponding two forms ${\cal O}^{(2)}$
are possible perturbations of the
topological Lagrangian.  The case of $p=q=1$ has particular significance,
since $H^1(X,T^{1,0})$ is the tangent space to the space of complex structures
on $X$, and the corresponding ${\cal O}^{(2)}$'s are just the changes
in the Lagrangian required by a change of complex structure.
(The explicit calculation showing this would be just analogous to the one
we will do presently for perturbations determined by elements of
$H^2(X,\wedge^2T^{1,0})$.)
So let
us discuss what happens when the complex structure of $X$ is changed.

The complex structure of $X$ is determined
by the $\bar\partial $ operator
$$\bar\partial=\sum_i\eta^{\bar i}{\partial\over\partial\phi^{\bar i}}.
\etag\yoro$$
(Mathematically, $\eta^{\bar i}$ would usually be written as the
$(0,1)$ form $d\phi^{\bar i}$.)  The transformation laws of the
$B$ model (for the fields $\phi,\eta,\theta$ from which the
basic observables are constructed)
are just the commutators with $\bar\partial$.
In \yoro, I have written the $\bar\partial$ operator acting on
$(0,q)$ forms (that is, functions of $\phi^I$ and $\eta^{\bar i}$),
but one can introduce the analogous $\bar\partial$ operator for
$(0,q)$ forms with values in any holomorphic bundle.  In our application,
the important holomorphic bundle is $\oplus_q\wedge^qT^{1,0}X$.
$(0,q)$ forms with values in this bundle are simply functions
of $\phi^I,\eta^{\bar i}$, and $\theta_j$.

If one makes a change in complex structure of $X$, the $\bar\partial$
operator changes.  To first order,
we get
$$\bar\partial\to \eta^{\bar i}\left({\partial\over\partial\phi^{\bar i}}
     +h_{\bar i}{}^j{\partial\over\partial\phi^j}
 -{\partial\over\partial\phi^k}h_{\bar i}{}^j\cdot\theta_j{\partial\over
\partial \theta_k}\right), \etag\yilop$$
where $h_{\bar i}{}^j$ is a cocycle representing an element of
$H^1(X,T^{1,0})$.
(The perturbed $\bar\partial$ operator, acting on functions or $(0,q)$
forms, may be more familiar; it is given by the same expression without
the $\theta\cdot \partial/\partial\theta $ term.  This term must be included
to give the
perturbed $\bar\partial$ operator acting on $(0,q)$ forms valued
in $\oplus_q\wedge^qT$.)
When the complex structure of $X$ is changed,
the transformation laws of the $B(X)$ model therefore also change;
indeed, taking the commutator with the perturbed $\bar\partial$ operator,
we find
$$\eqalign{\delta\phi^i & =i\alpha \eta^{\bar i}h_{\bar i}{}^j,\cr
\delta\theta_j
&  =-i\alpha\eta^{\bar i}\partial_jh_{\bar i}{}^s\theta_s,\cr}\etag\hopow$$
which replace $\delta \phi^i=\delta\theta_j=0$ in the unperturbed theory.
The non-zero transformation law of $\theta_j$ is not so essential
in the following sense: it reflects the change in $T^{1,0}$ (of which
$\theta$ is a section) under a change in the complex structure of $X$,
and it can be transformed away by rotating the $\theta^i$ to a basis
appropriate to the new complex structure.  The non-zero transformation
law of $\phi^i$ is unavoidable, as it is a basic expression of the change
in complex structure.

\subsubsec{A Non-Classical Case}

So even in a ``classical'' case, where one is just perturbing the
complex structure of $X$, deformations of the $B$ model require
a change in the transformation laws of the basic fields.
One must expect this to be true also for other, less classical
deformations.

As a typical example, let $\alpha$ be a cocycle representing
an element of $H^2(X,\wedge^2T^{1,0})$.
The corresponding BRST invariant
operator-valued zero form is
$${\cal O}^{(0)}=\alpha_{\overline i_1\bar i_2}{}^{j_1j_2}\eta^{\bar i_1}
\eta^{\bar i_2}\theta_{j_1}\theta_{j_2}              . \etag\corrzer$$
We now wish to write $d{\cal O}^{(0)}=\{Q,{\cal O}^{(1)}\}$, for some
${\cal O}^{(1)}$.  In constrast to the $A$ model, one finds immediately that
(i) this is only true modulo terms that vanish by the equations of motion;
(ii) the calculations involved are rather painful.
The second point is almost inevitable (in the absence of a powerful
computational framework) given the first; and the first point
is related, as we will see, to the fact that under perturbation of
the Lagrangian, the transformation laws of the fields change.

Eventually one finds that
$$ d{\cal O}^{(0)}= \{Q,{\cal O}^{(1)}\} +  G,\etag\finalfo$$
where
$$\eqalign{
{\cal O}^{(1)}= &i\rho^iD_i\alpha_{\bar i_1\bar i_2}{}^{j_1j_2}\eta^{\bar i_1}
\eta^{\bar i_2}\theta_{j_1}\theta_{j_2}+2d\phi^{\bar i_1}\alpha_{\bar i_1
\bar i_2}{}^{j_1j_2}\eta^{\bar i_2}\theta_{j_1}\theta_{j_2}
\cr &-2\alpha_{\bar i_1\bar i_2}{}^{j_1j_2}\eta^{\bar i_1}\eta^{\bar i_2}
\theta_{j_1}g_{j_2\overline k} \star dX^{\bar k}.\cr} \etag\pilofo$$
The $\star$ here is the Hodge star operator,
and
$$G=2\alpha_{\bar i_1\bar i_2}{}^{j_1j_2}\eta^{\bar i_1}\eta^{\bar i_2}
\theta_{j_1}Z_{j_2}, \etag\molfo$$
where
$$Z_j=D\theta_j-i\rho^m\eta^{\bar j}\theta_sR_{j{\bar j}m}{}^s
- g_{j\bar j}\star D\eta^{\bar j} \etag\polfo$$
vanishes by the $\rho$ equation of motion.

The next step is to solve the equations
$$d{\cal O}^{(1)}=\{Q,{\cal O}^{(2)}\}+\sum_A{\delta L\over\delta \Phi_A}
        \cdot \zeta_A.    \etag\piloz$$
Here $\Phi_A$ are all the fields of the theory
($\phi,\eta,\theta,\rho$).  Moreover, $\delta L/\delta\Phi_A$ are the equations
   of
motion of the theory, so any expression that vanishes by the equations of
motion is of the form $\sum_A\delta L/\delta \Phi_Z\cdot \zeta^A$
for some $\zeta^A$.  \piloz\ means that in forming the topological family,
the generalized Lagrangian
$$\widetilde L=L+t\int_\Sigma{\cal O}^{(2)}\etag\hoffo$$
is not invariant under the original BRST transformations, but is invariant
(to this order; that is,
up to terms of order $t^2$) under
$$\widetilde \delta \Phi_A=\delta\Phi_A+t\zeta_A.\etag\yoffo$$
This shows how the modifications of the transformation laws,
which we anticipate from our preliminary discussion of the role of
$H^1(X,T^{1,0})$, depend upon having non-zero $\zeta_A$.

The computation involved in finding
${\cal O}^{(2)}$ in \piloz\ (which is guaranteed to exist
by the general discussion at the beginning of this section) is very
complicated,
and would be unilluminating if done without powerful computational methods,
such as a superspace formulation.
It is much easier to determine the $\zeta_A$.
The $\zeta_A$ can be determined to eliminate terms in
$d{\cal O}^{(1)}$ that do not appear in any expression of the general
form $\{Q,{\cal O}^{(2)}\}$.  I will just state the results.
First of all, one finds that
$\zeta_\eta=0$.  This is in fact inevitable, even without computation,
to preserve the fact that $Q^2=0$ (when acting on $\phi^{\bar i}$).
The important novelty, compared to the derivation of
equation \finalfo,
is that $\zeta_\theta$ and $\zeta_\phi$ are non-zero.
In fact,
$$\zeta_{\phi^j}=-2i\alpha_{\bar i_1\bar i_2}{}^{jk}\eta^{\bar i_1}
\eta^{\bar i_2}\theta_k \etag\torpop$$
and
$$\zeta_{\theta_i}=iD_i\alpha_{\bar i_1\bar i_2}{}^{j_1j_2}\eta^{\bar i_1}
\eta^{\bar i_2}\theta_{j_1}\theta_{j_2}. \etag\morpop$$

Including the terms necessitated by \torpop\ and \morpop, the
BRST transformation laws of the topological family (or rather,
the one parameter subfamily determined by the particular ${\cal O}^{(2)}$
considered here) are
$$\eqalign{
\delta\phi^{\overline i} & = i\alpha\eta^{\bar i}   \cr
 \delta\eta^{\bar i} & = 0       \cr
 \delta\phi^j & = -2it\alpha_{\bar i_1\bar i_2}{}^{jk}\eta^{\bar i_1}
\eta^{\bar i_2}\theta_k \cr
\delta \theta^i & =itD_i\alpha_{\bar i_1\bar i_2}{}^{j_1j_2}\eta^{\bar i_1}
\eta^{\bar i_2}\theta_{j_1}\theta_{j_2}.
\cr} \etag\gorfo$$

What sort of perturbed $\bar\partial$ operator will generate such
transformations?  Evidently, we need
$$\overline D=\eta^{\bar i}{\partial\over\partial\phi^{\bar i}}
-2it\eta^{\bar i_1}\eta^{\bar i_2}\alpha_{\bar i_1\bar i_2}{}^{j_1j_2}
\theta_{j_1}{\partial\over\partial \phi^{j_2}}+it\eta^{\bar i_1}\eta^{\bar i_2}
D_k\alpha_{\bar i_1\bar i_2}{}^{j_1j_2}\theta_{j_1}\theta_{j_2}
{\partial\over\partial \theta_k}.     \etag\polyo$$

\subsubsec{Interpretation}

{}From this sample computation, it is possible to guess the general
structure, as we will now indicate.
Like the original $\bar\partial$ operator, $\bar D$ is a first
order differential operator, which acts
on functions of $\phi,\eta,\theta$.
As it is the BRST operator of the perturbed model, it
obeys $\bar D^2=0$ (after adding
terms of order $t^2$ and higher, which we have
not analyzed).\foot{$\bar D$ is a symmetry of the perturbed model,
so $\bar D^2$ is likewise a symmetry, which moreover is bosonic and of ghost
number two.
The perturbed or unperturbed model has no such symmetries, so it must
be that $\bar D^2=0$.}
However, because of the terms of third and higher
order in fermions, it is definitely not a classical $\bar\partial$ operator.

Let ${\cal M}$ be the moduli space of complex structures on $X$,
modulo diffeomorphism.  ${\cal M}$ parametrizes the $B(X)$ models
as we constructed them originally in \S4.
The
``topological family'' of theories (with Lagrangian $L\to L+\sum_a
t_a\int {\cal O}_a^{(2)}$) is a more general family of theories parametrized
by a thickened moduli space ${\cal N}$ which contains ${\cal M}$ as a
subspace; the tangent space to ${\cal N}$ at
${\cal M}\subset {\cal N}$ is $T{\cal N}|_{{\cal M}}=\oplus_{p,q=0}
^d H^p(X,\wedge^qT^{1,0}X)$.

The moduli space ${\cal M}$ of complex structures on $X$
is the space of standard $\bar\partial$ operators
(first order operators whose leading symbol is linear in $\eta^{\bar i}$
and independent of $\theta$), which obey $\bar\partial^2=0$,
modulo diffeomorphisms of the $\phi^I$.
The enriched moduli space ${\cal N}$
of topological models is apparently, in view of \polyo, the space
of general $\bar D$ operators -- first order operators on the same
space, but with the leading symbol allowed to have a general dependence
on $\phi,\eta,\theta$
-- obeying $\bar D^2=0$,
modulo diffeomorphisms of $\phi^I,\,\eta^{\overline i},\,\theta_i$.
(Thus, $\bar\partial$ operators are classified up to ordinary diffeomorphisms
of $\phi$ only, but $\bar D$ operators are classified up to diffeomorphisms
of the $\phi,\eta,\theta$ supermanifold.)
The replacement of classical $\bar\partial$
operators by the more general $\bar D$ operators has some of the flavor
of string theory, where classical geometry is generalized in a somewhat
analogous, but far more drastic, way.

I will call ${\cal N}$ the extended moduli space.
The analogous extended moduli space in the $A$ model
is just $\oplus_{n=0}^{2d} H^n(X,\IC)$.
The study of the extended moduli space  is probably the proper framework for
understanding the mirror map between the $A$ and $B$ models, and so is
potentially rewarding.

\subsec{The Mirror Map}

The weakest link in existing studies of the consequences of mirror
symmetry is the construction of the mirror map between the mirror
moduli spaces.
Given, in other words, a mirror pair $X,Y$, one would like to identify
the mirror map between the $A(X)$ moduli space and the $B(Y)$ moduli space.
Understanding this map is essential for extracting the consequences of mirror
symmetry.
The discussion in [\candelas] involved
an elegant and very successful but not fully understood ansatz.

To construct the mirror map, it suffices to identify some sufficiently
rich class of observables that can be computed both in the $A(X)$ model
and in the $B(Y)$ model.  I do not know how to do this, but will make
a few comments.  Since the $B(Y)$ model reduces to classical
algebraic geometry, ``everything'' is computable in the $B(Y)$ model.
The $A(X)$ model is another story; observables in the $A(X)$ model
depend on complicated instanton sums and so in general are not really
calculable (except perhaps via mirror symmetry which requires already
knowing the mirror map).

The association ${\cal O}^{(0)}\to {\cal O}^{(2)}$
that we have described above gives a natural identification between
the tangent space to the extended moduli space (which is the space
of ${\cal O}^{(2)}$'s) and the BRST cohomology classes of local
operators (the ${\cal O}^{(0)}$'s).  The two point function
in genus zero
$$\eta_{ab}=\langle {\cal O}_a^{(0)}{\cal O}_b^{(0)}\rangle
     \etag\jhj$$
therefore defines a metric on the extended moduli space.
\foot{I should stress that this ``topological'' metric
in no way coincides with the Zamolodchikov metric on the moduli
space -- which is much more difficult to study and involves what has
been called ``topological anti-topological fusion'' [\vvvafa].}
(It is essential here to work with the extended moduli space ${\cal N}$;
the induced metric on the ordinary moduli space ${\cal M}\subset {\cal N}$
is typically degenerate or even zero.)  Moreover, it can be shown
that this metric is flat.  For topological field theories (like the $A$ and
$B$ models for Calabi-Yau manifolds) that arise
by twisting conformal field theories, this was shown in [\dijkgraaf].
The main ingredients in the argument are the $\IC^*$ action on a genus
zero surface with two marked points, and the fact that
for the particular BRST invariant local operator ${\cal O}^{(0)}=1$,
the corresponding two form is ${\cal O}^{(2)}=0$.
For the $A$ model, I will give another proof of flatness presently
(by showing that the metric has no instanton corrections); this proof
uses the $\IC^*$ action in a different way, and does not require the
Calabi-Yau condition.

For the $A(X)$ model, the extended moduli space is $\sum_{n=0}^{2d} H^2(X,
\IC)$, and the flat metric
is simply the metric on this vector space given
by Poincar\'e duality; there are no instanton corrections,
as we will show shortly.  For topological field
theories obtained by twisting Landau-Ginzberg models, the
metric
was computed by Vafa [\vvafa] up to a conformal factor; that the metric
that so arises is flat (after a proper choice of the conformal factor,
which has not yet been analyzed in the quantum field theory)
is part of the rather deep study of singularities by
K. Saito [\saito], as was explained by
Blok and Varchenko [\blok].  For the $B(Y)$ model, the extended moduli
space was roughly described above, but the metric is not yet understood.
It seems to be very hard to find any observables of the $A(X)$ model
except the metric (and the related ``exponential map'' noted below) that are
effectively computable, without instanton corrections.  So understanding
the metric on the extended moduli space of the $B(Y)$ model would
appear to be a promising way to understand the mirror map.
(If the metric were known for both $A(X)$ and $B(Y)$,
the mirror map would be uniquely determined
up to an isometry; isometries depend on
finitely many constants, which one might determine by matching a few
coefficients at infinity.)

\subsubsec{Vanishing Of Instanton Corrections To The Metric}

It remains to show that instantons do not contribute to
the metric of the $A(X)$ model.  The proof
is essentially dimension counting, taking account of the $\IC^*$
action on a genus zero surface with two marked points.
Let $X$ be a complex manifold, not necessarily Calabi-Yau, and $\Sigma$ a
Riemann
surface of genus zero.
Consider a component ${\cal U}$ of the moduli space of holomorphic maps
$\Phi:\Sigma\to X$ of
virtual dimension $w$.
Let $H$ and
$H'$ be submanifolds (or homology cycles) of codimension $q$ and $q'$
with $q+q'=w$, and let $P$ and $P'$ be two points in
$\Sigma$.  Let ${\cal U}'$ be the subspace of ${\cal U}$ parameterizing
$\Phi$'s with $\Phi(P)\in H$ and
$\Phi(P')\in H'$.  If ${\cal U}'$ is empty, the contribution of this homotopy
class
to $\langle {\cal O}_H^{(0)}{\cal O}_{H'}^{(0)}\rangle$ is zero.
As the virtual dimension of ${\cal U}'$ is zero, it appears superficially
that ${\cal U}'$ need not vanish generically.

However, let us take account of the $\IC^*$ action on $\Sigma$
fixing $P$ and $P'$.   Let $\widetilde{\cal U}={\cal U}'/\IC^*$.
${\cal U}'$ must be empty if $\widetilde{\cal U}$ is.
One can think of $\widetilde{\cal U}$ as the moduli space of
rational curves $C\in X$ that intersect both $H$ and $H'$
(without specifying any parameterization of $C$ or map from $\Sigma$
to $C$).
The virtual dimension of $\widetilde {\cal U}$ is $-2$.
Hence, $\widetilde {\cal U}$ is ``generically'' empty, and will really
be empty at worst
after making a generic nonintegrable deformation of the almost
complex structure of $X$.  So the instanton corrections to the metric vanish.

As a very concrete example of this counting,
let $X$ be a particular Calabi-Yau manifold, a quintic hypersurface
in ${\bf P}^4$.  The virtual dimension
of ${\cal U}$ is six (regardless of the homotopy class of the map considered).
Let $H$, $H'$ be two
homology cycles in $X$ the sum of whose codimensions is six.
So one of them, say $H$, has codimension
at least three.  With this particular $X$, even for generic
integrable complex structures, it
is believed that the number  of rational curves
of given positive degree in $X$ is finite;
if so, the union $Z$ of these curves has codimension four.  Since $4+3>6$, $H$
can be perturbed slightly so as not to intersect $Z$; hence ${\cal U}'$
is empty, and
the instanton contribution to the metric
vanishes.

\subsubsec{The Exponential Map}

The description of the topological family by a family of Lagrangians
$$\widetilde L=L+\sum_at_a\int_\Sigma {\cal O}_a^{(2)}
           \etag\upo$$
shows that once we pick a base Lagrangian $L$, corresponding to a base
point $P\in {\cal N}$, there is a natural
linear structure on ${\cal N}$.  This might be described as
an exponential map from the tangent space to ${\cal N}$ at $P$
to ${\cal N}$.  In the case of the $A$ model, this structure
corresponds to the linear structure on $\oplus_n H^n(X,\IC)$.
In the case of the $B$ model, it is not presently understood.
Understanding the exponential map should be roughly similar to understanding
the metric on the moduli space.  It is puzzling, in particular, that in
the $A$ model, the linear structure on the moduli space does not seem to depend
on the choice of base point, while in the $B$ model such a dependence
seems almost inevitable.

\listrefs\bye